\documentclass{article}
\usepackage{amsmath,graphicx}
\usepackage[preprint]{spconf}

\title{Soft Dynamic Time Warping for Multi-Pitch Estimation and Beyond}
\name{Michael Krause, Christof Weiß, Meinard~M{\"u}ller\thanks{This work was supported by the German Research Foundation (DFG MU 2686/7-2). The authors are with the International Audio Laboratories Erlangen, a joint institution of the Friedrich-Alexander-Universit\"at Erlangen-N\"urnberg (FAU) and Fraunhofer Institute for Integrated Circuits IIS. The authors gratefully acknowledge the compute resources and support provided by the Erlangen Regional Computing Center (RRZE).}}
\address{International Audio Laboratories Erlangen}
\usepackage{xcolor}
\usepackage{booktabs}
\usepackage{multirow}
\usepackage{units}
\usepackage{subcaption}
\usepackage{tikz}
\usepackage{mathtools}
\usepackage{amsfonts}
\usepackage{cite}
\newcommand{\ciec}{, i.\,e.,\ }
\newcommand{\iec}{i.\,e.,\ }

\newcommand{\cegc}{, e.\,g.,\ }

\newcommand{\Sect}[1]{Section~\ref{#1}}
\newcommand{\Fig}[1]{Figure~\ref{#1}}
\newcommand{\Table}[1]{Table~\ref{#1}}

\DeclareMathOperator{\N}{{\mathbb N}}

\DeclareMathOperator{\R}{{\mathbb R}}

\DeclarePairedDelimiter\abs{\lvert}{\rvert}%

\newcommand\copyrighttext{%
	\footnotesize \textcopyright 2023 IEEE. Personal use of this material is permitted.  Permission from IEEE must be obtained for all other uses, in any current or future media, including reprinting/republishing this material for advertising or promotional purposes, creating new collective works, for resale or redistribution to servers or lists, or reuse of any copyrighted component of this work in other works.}

\copyrightnotice{
	\begin{tikzpicture}[remember picture,overlay]
		\node[anchor=south,yshift=10pt] at (current page.south) {\fbox{\parbox{\dimexpr\textwidth-\fboxsep-\fboxrule\relax}{\copyrighttext}}};
	\end{tikzpicture}
}

\begin{document}
\ninept
\maketitle
\begin{abstract}
Many tasks in music information retrieval (MIR) involve weakly aligned data, where exact temporal correspondences are unknown. The connectionist temporal classification (CTC) loss is a standard technique to learn feature representations based on weakly aligned training data. However, CTC is limited to discrete-valued target sequences and can be difficult to extend to multi-label problems. In this article, we show how soft dynamic time warping (SoftDTW), a differentiable variant of classical DTW, can be used as an alternative to CTC. Using multi-pitch estimation as an example scenario, we show that SoftDTW yields results on par with a state-of-the-art multi-label extension of CTC. In addition to being more elegant in terms of its algorithmic formulation, SoftDTW naturally extends to real-valued target sequences.
\end{abstract}
\section{Introduction}
\label{sec:intro}
\begin{figure}
	\centering
	\includegraphics[width=0.9\columnwidth]{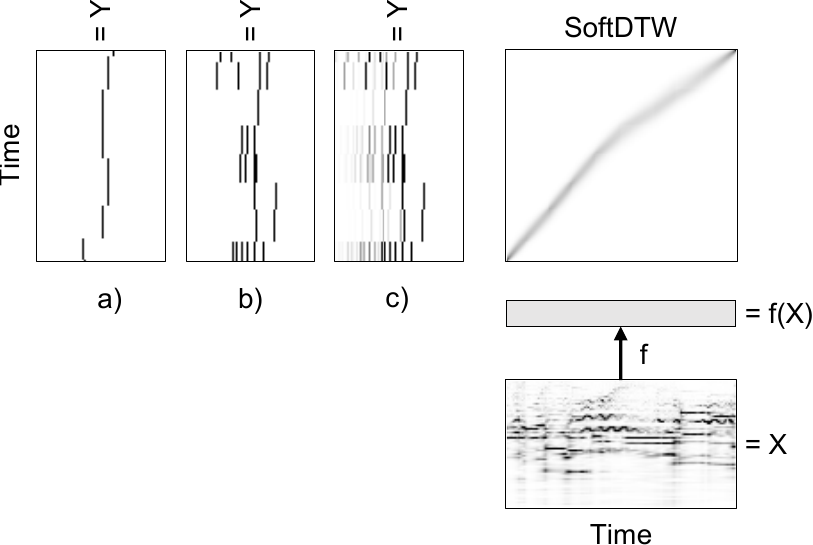}
	\caption{Illustration of SoftDTW for aligning a learned feature sequence $f(X)$ and a target sequence Y, where one may consider (a) single-label, (b) multi-label, or (c) real-valued targets.}
	\label{fig:teaser}
	\vspace*{-10pt}
\end{figure}

Many applications in music information retrieval (MIR) require alignments between  sequences of music data. Often, the sequences given are only weakly aligned. For example, in audio-to-score transcription, pairs of audio and score excerpts are easy to find but exact temporal correspondences between these pairs are hard to establish \cite{WeissP21_MultiPitchMCTC_WASPAA}. Furthermore, music data sequences may involve different levels of complexity. For instance, given a single-instrument monophonic music recording, monophonic pitch estimation \cite{BoschBSG16_MelodyExtraction_ISMIR} aims at finding a single pitch value per time step (see also \Fig{fig:teaser}a). Other scenarios with discrete, single-label targets include lyrics transcription or lyrics alignment for songs with a single singer \cite{WangKNSY04_syncLyrics_ACMMM,Schulze-ForsterDRB21_LyricsDTW_TASLP}. More complex sequences appear in multi-pitch estimation (MPE), where multiple pitches may be active simultaneously (\Fig{fig:teaser}b). Finally, some scenarios involve alignment between real-valued sequences (\Fig{fig:teaser}c)\cegc audio--audio synchronization \cite{DixonW05_MATCH_ISMIR,EwertMG09_HighResAudioSync_ICASSP} or multi-modal alignment problems such as synchronizing dance videos with music \cite{TsuchidaFHG19_DanceProcessing_ISMIR}.

The connectionist temporal classification (CTC) \cite{GravesFGS06_CTCLoss_ICML} loss, a fully differentiable loss function initially developed for speech recognition, is commonly used for learning features from weakly aligned data when the targets are sequences over a finite alphabet of labels. Recently, CTC was extended to handle multi-label learning problems \cite{WigingtonPC19_MultiLabelCTC_ICDAR}, where the main idea was to locally transform the multi-label into the single-label case. However, in addition to its complicated algorithmic formulation, this approach is unsuitable for target sequences that do not originate from a discrete vocabulary.

A common technique used in MIR for finding an optimal alignment between weakly aligned sequences is dynamic time warping (DTW) in combination with hand-crafted features  \cite{Mueller21_FMP_SPRINGER}. Such a pipeline can provide good alignment results for tasks like audio--audio synchronization \cite{EwertMG09_HighResAudioSync_ICASSP}, but the standard DTW-based cost function is not fully differentiable, which prevents its use in an end-to-end deep learning context. To resolve this issue, Cuturi and Blondel \cite{CuturiB17_SoftDTW_ICML} proposed a differentiable variant of DTW, called SoftDTW, that approximates the original DTW cost. In recent work, SoftDTW and related techniques have been successfully used in computer vision applications such as action alignment \cite{HadjiDJ21_SmoothDTW_CVPR,ChangHS0N19_D3TW_CVPR}. 
To our knowledge, the only prior work applying SoftDTW in an MIR context is by Agrawal et al. \cite{AgrawalWD_ConvolutionalScoreAudioSync_SPL}.

Our contributions are as follows: We demonstrate the use of SoftDTW for MPE. In particular, we show that SoftDTW performs on par with a multi-label extension of CTC, while being conceptually simpler. Furthermore, we show that the SoftDTW approach naturally generalizes to real-valued target sequences, as illustrated in \Fig{fig:teaser}, making it applicable for a wide range of alignment tasks. 

The remainder of the paper is structured as follows: In \Sect{sec:mpe}, we review the current state of the art for multi-pitch estimation from weakly aligned data with CTC. In \Sect{sec:softdtw}, we formalize SoftDTW for general sequences and, in \Sect{sec:softdtw-mpe}, apply it for MPE. \Sect{sec:real} demonstrates the potential of SoftDTW for learning with real-valued targets. Finally, \Sect{sec:conclusion} concludes the paper with an outlook towards future applications.

\section{Weakly Aligned Training for MPE}
\label{sec:mpe}
In recent years, automated music transcription has become a central topic in MIR research, with deep learning techniques achieving state-of-the-art results
\cite{HawthorneESRSRE18_OnsetsFrames_ISMIR,KelzDKBAW16_FramewisePianoTrans_ISMIR,CheukLBH21_OnsetsFramesAttention_IJCNN}. We here focus on MPE as a sub-problem of automated music transcription, where the goal is to transform an input music recording $X$ into a piano-roll representation $Y$ of pitches played. In particular, multiple pitches may be active at the same time. Most learning-based approaches for MPE require strongly aligned data for training\ciec pitches are annotated for each audio frame of the input recording. Since annotating data in such a frame-wise fashion is very time consuming, most MPE datasets have been generated (semi-)automatically\cegc by using MIDI pianos or by applying score--audio synchronization techniques (which may introduce labeling errors). Techniques that allow learning from pairs of $X$ and $Y$ that are not temporally aligned are therefore highly desirable.

As discussed in the introduction, a common technique for dealing with weakly aligned learning problems is CTC \cite{GravesFGS06_CTCLoss_ICML}. Here, the target sequences $Y$ consist of symbols from a discrete alphabet $L$, including a special blank symbol necessary for distinguishing repetitions of symbols. For each frame in the input sequence $X$, a neural network outputs a probability distribution over $L$. The CTC loss then corresponds to the likelihood of $Y$ given these network outputs, taking into account all possible alignments between $X$ and $Y$. Note that CTC is agnostic about the durations of symbols in $Y$\ciec even if information about symbol durations is available, CTC is unable to exploit this for alignment. An efficient dynamic programming algorithm for computing the CTC loss exists (with time complexity $\mathcal{O}(\abs{L}^2\cdot N)$, where $N$ is the length of $X$), but it requires special care in handling the blank symbol \cite{GravesFGS06_CTCLoss_ICML}.

A naive extension of CTC towards multi-label target sequences would introduce unique network outputs for all possible symbol combinations, which leads to a combinatorial explosion. Instead, the authors in \cite{WigingtonPC19_MultiLabelCTC_ICDAR} propose to locally reduce the multi-label to the single-label case by only considering those symbol combinations that occur within a single training batch (called multi-label CTC\ciec MCTC). This defines a ``batch-dependent alphabet,'' avoiding the combinatorial explosion. The technical details of this process are tricky and special care needs to be taken for handling the blank symbol. In \cite{WeissP21_MultiPitchMCTC_WASPAA}, this idea is adapted for MPE by considering pitches as symbols and multi-pitch annotations as combinations of symbols. This formulation allows them to train networks for MPE on pairs of $X$ and $Y$ that are only weakly aligned\cegc where $X$ is a music recording and $Y$ is a MIDI representation derived from the corresponding score. In this paper, using MPE from \cite{WeissP21_MultiPitchMCTC_WASPAA} as an example application, we show how the technically intricate MCTC can be replaced by a conceptually more elegant SoftDTW approach. SoftDTW does not involve the need for a blank symbol, which may be well-motivated in text applications but can be unnatural in MIR problems such as MPE.

\section{Soft Dynamic Time Warping}
\label{sec:softdtw}
The objective of DTW is to find an optimal temporal alignment between two sequences. SoftDTW \cite{CuturiB17_SoftDTW_ICML} is a differentiable approximation of DTW that allows for propagating gradients through the alignment procedure, making SoftDTW applicable for deep learning. Like classical DTW, SoftDTW admits an efficient dynamic programming (DP) recursion for computing the optimal alignment cost. Furthermore, there also exists a DP-algorithm for efficiently computing the gradient of that cost. In this section, we briefly summarize the problem statement and DP recursion of SoftDTW for general sequences. We then apply this to our music scenarios in later sections.

Consider two sequences $X=\left(x_1, x_2, \dots, x_N\right)$ and $Y=\left(y_1, y_2, \dots, y_M\right)$ of lengths $N, M\in\N$ with elements coming from some feature spaces $\mathcal{F}_1, \mathcal{F}_2$ (\iec  $x_n\in\mathcal{F}_1,y_m\in\mathcal{F}_2$ for all $n\in\left[1:N\right], m\in\left[1:M\right]$). Given some differentiable cost function $c: \mathcal{F}_1\times\mathcal{F}_2\rightarrow\R$ defined on these feature spaces, we can construct a matrix $C\in\R^{N\times M}$ of local costs where each entry
\[C(n,m)=c(x_n, y_m)\]
contains the cost of \textit{locally} aligning $x_n$ with $y_m$. To determine an optimal global alignment\footnote{Subject to some constraints, namely, the first and last elements of both sequences are aligned to each other (boundary constraint), no element is skipped (step-size constraint), and the alignment is monotonous (monotonicity constraint).} between the sequences $X$ and $Y$ one computes an accumulated cost matrix $D^\gamma\in\R^{N\times M}$ using the recursion
\begin{align*}
D^\gamma(1,1)&=C(1,1),\\
D^\gamma(1,m) &= \sum_{k=1}^{m}C(1,k), \text{ for } m\in\left[1:M\right],\\
D^\gamma(n,1) &= \sum_{k=1}^{n}C(k,1), \text{ for } n\in\left[1:N\right],\\
D^\gamma(n,m)&=C(n,m) + \mu^\gamma(\{D^\gamma(n-1, m-1), \\
&D^\gamma(n-1, m), D^\gamma(n, m-1)\}),
\end{align*}
for $n\in\left[2:N\right], m\in\left[2:M\right]$.
Here, $\mu^\gamma$ refers to a differentiable approximation of the minimum function given by
\[\mu^\gamma(S)=-\gamma\log\sum_{s\in S}\exp\left(-\frac{s}{\gamma}\right),\]
where $S$ is some finite set of real numbers and $\gamma\in\R^{>0}$ is a temperature parameter that determines the ``softness" of the approximation.
One can show that $\mu^\gamma$ is a lower bound of the minimum function \cite{HadjiDJ21_SmoothDTW_CVPR} and converges towards the true minimum for $\gamma\rightarrow 0$. As a consequence, $D^\gamma$ becomes the accumulated cost matrix from classical DTW for $\gamma\rightarrow 0$. Thus, SoftDTW becomes DTW in the limit case.

After evaluating the SoftDTW recursion, the entry $\text{DTW}^\gamma(C)=D^\gamma(N,M)$ contains the approximate minimal cost of aligning the sequences $X$ and $Y$, given the local costs $C$. A similar recursion exists for computing the gradient of $\text{DTW}^\gamma(C)$ with regard to any matrix coefficient $C(n,m)$ for $n\in\left[1:N\right]$ and $m\in\left[1:M\right]$ \cite[Algorithm 2]{CuturiB17_SoftDTW_ICML}. The time and space complexity of the SoftDTW recursion as well as the gradient computation is both $\mathcal{O}(N\cdot M)$, which is sufficiently fast for use in deep learning.

Note that SoftDTW requires no prior knowledge of the alignment between $X$ and $Y$, which enables the use of $\text{DTW}^\gamma(C)$ as a loss function for learning problems with weakly aligned data. Furthermore, $X$ and $Y$ can come from arbitrary feature spaces, as long as an appropriate cost function $c$ can be defined. 

\section{Application to Multi-Pitch Estimation}
\label{sec:softdtw-mpe}
We now apply SoftDTW to multi-pitch estimation. For a given piece of music, the sequence $X$ corresponds to some representation of an input recording, while $Y$ corresponds to a multi-hot encoding of pitches played. Note that $Y$ does not need to be temporally aligned with $X$ and could arise\cegc from a score representation of the musical piece. An element $y_m$ of the sequence $Y$ is encoded as a vector $y_m\in\{0,1\}^{72}$ and the entries of $y_m$ correspond to the $72$ pitches from C1 to B6. In our experiments, rather than directly aligning $Y$ with some fixed representation $X$, we use a neural network $f$ that takes $X$ as input and outputs a feature vector per frame in $X$. Thus, we obtain a sequence $f(X)=(z_1,\dots,z_N)$ with the same length $N$ as $X$. We construct $f$ such that $z_n\in\R^{72}$ for the elements $z_n$ of $f(X)$. Thus, both sequences $Y$ and $f(X)$ contain elements from the features space $\mathcal{F}_1=\mathcal{F}_2=\R^{72}$. We then align $f(X)$ and $Y$, as illustrated in \Fig{fig:teaser}. 

To our knowledge, SoftDTW has not previously been used for MPE and is seldom explored in MIR. The authors in \cite{Schulze-ForsterDRB21_LyricsDTW_TASLP} used the classical, non-differentiable DTW recursion inside an attention mechanism for lyrics alignment, which led to training instabilities. The work by Agrawal et al. \cite{AgrawalWD_ConvolutionalScoreAudioSync_SPL} constitutes the first use of SoftDTW for an MIR application. They successfully employ a variant of SoftDTW to train a system for score-audio synchronization. In their scenario, SoftDTW is applied to discrete-valued, one-dimensional, and strongly aligned sequences. In contrast, we employ SoftDTW for multi-dimensional sequences in weakly aligned settings.\vspace*{-6pt}

\subsection{Implementation Details and Evaluation Metrics}
\label{sec:softdtw-mpe:implementation-details}
Since the focus of our work is on evaluating the efficacy of SoftDTW for MIR tasks and in order to maintain comparability with the results presented in \cite{WeissP21_MultiPitchMCTC_WASPAA}, we adopt the same training setup and network architecture. Thus, we use harmonic CQT (HCQT, \cite{BittnerMSLB17_DeepSalience_ISMIR}) excerpts of roughly ten second lengths as input and pass them through a five-layer convolutional neural network to obtain a sequence of per-frame representations $f(X)$ (see \cite{WeissP21_MultiPitchMCTC_WASPAA} for details on the network architecture and HCQT representation).

We train our networks by minimizing the soft alignment cost $\text{DTW}^\gamma(C)$.\footnote{Note that we normalize $\text{DTW}^\gamma(C)$ by its value for the first training batch. Thus, the loss is exactly $1$ for the first batch and its value range remains similar across training configurations, regardless of the sequence lengths $N$ and $M$ or other factors.} 
In all experiments, we use the squared Euclidean distance for $c$ and set $\gamma=10.0$. We did not see improvements for alternative choices of $c$ and obtained similar results for a wide range of values for $\gamma\in\left[0.5, 20.0\right]$. Furthermore, we use a fast GPU implementation of the SoftDTW recursion and gradient computation which was implemented in \cite{MaghoumiTL_SoftDTWPytorch_IUI}.

To compare network predictions with the strongly aligned pitch annotations of the test sets, we use common evaluation measures for MPE, including cosine similarity between predictions and annotations (CS), area under the precision-recall curve (also called average precision, AP), as well as F-measure and accuracy (Acc., introduced in \cite{PolinerE07_PolyphonicPiano_EURASIP}) at a threshold of 0.4 (which is a common choice in MPE systems, see also \cite{ThickstunHK17_MusicNet_ICLR}).\vspace*{-5pt}

\subsection{Comparison with MCTC}
\label{sec:softdtw-mpe:results1}
\begin{figure}
	\centering
	\includegraphics[scale=0.92]{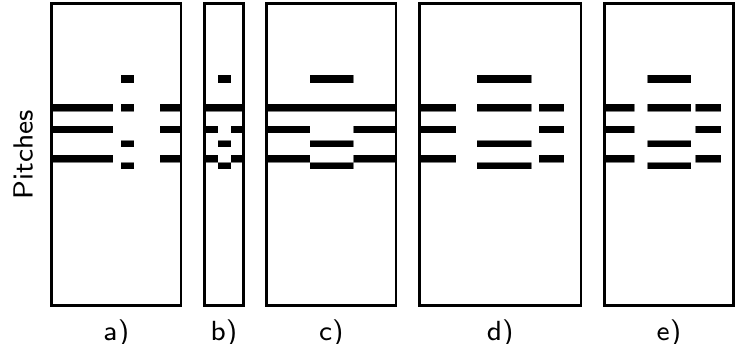}
	\caption{(a) Strongly aligned pitch annotations for an audio excerpt, (b) Annotations without note durations (as used by MCTC), (c) Annotations without note durations, stretched to excerpt length, (d) Score representation, not aligned to the audio excerpt, (e) Score representation, stretched to excerpt length}
	\label{fig:label-types}
	\vspace*{-5pt}
\end{figure}

We begin by comparing our results with the main results reported in \cite{WeissP21_MultiPitchMCTC_WASPAA}, which are obtained on the Schubert Winterreise Dataset (SWD) \cite{WeissZAMKVG21_WinterreiseDataset_ACM-JOCCH}. SWD provides strongly aligned annotations for all recordings. Due to this, one can consider a baseline trained on the aligned annotations with a per-frame cross-entropy loss (CE). The first line of \Table{tab:results:schubert1} shows results for such an optimistic baseline (reprinted from \cite{WeissP21_MultiPitchMCTC_WASPAA}), which yields an F-measure of $0.70$ and $\text{AP}=0.764$. To train a network using MCTC instead, one must remove all information about note durations from the label sequence $Y$ (see \Fig{fig:label-types}b). The results obtained this way are just slightly lower at $\text{AP}=0.734$, even though only weakly aligned labels are used. When performing the same experiment using SoftDTW (denoted by $\mathrm{SoftDTW}_\mathrm{W1}$), we obtain much weaker results with an F-measure of $0.00$ and $\text{AP}=0.297$.\footnote{Note that the F-measure and Accuracy scores can be improved to $0.32$ and $0.20$, respectively, by choosing a more suitable detection threshold. Still, these scores are notably worse compared to the results for MCTC.} In this experiment, the label sequence $Y$ may be significantly shorter than the learned sequence $f(X)$.\footnote{A large discrepancy in sequence lengths is well known to cause problems for classical DTW. Further investigation is needed to understand how this affects the training process with SoftDTW.} We repeat the experiment by temporally stretching the sequence $Y$ to match the number of frames in $f(X)$ (illustrated in \Fig{fig:label-types}c). When applying SoftDTW together with this trick (denoted by $\mathrm{SoftDTW}_\mathrm{W2}$), results are again very similar to MCTC ($\text{AP}=0.737$). Thus, SoftDTW may be used to replace MCTC in this scenario.

\begin{table}[t]
	\centering
	{\footnotesize
		\setlength{\tabcolsep}{4.5pt}
		\begin{tabular}{lcccr}			
			\toprule
			\textbf{Scenario} & \textbf{F-measure} & \textbf{CS} & \textbf{AP} & \textbf{Acc.} \\ \midrule
			CE \cite{WeissP21_MultiPitchMCTC_WASPAA} & 0.70 & 0.759 & 0.764 & 0.546 \\
			MCTC \cite{WeissP21_MultiPitchMCTC_WASPAA} & 0.69 & 0.744 & 0.734 & 0.532 \\\midrule
			$\mathrm{SoftDTW}_\mathrm{W1}$ & 0.00 & 0.465 & 0.297 & 0.002 \\
			$\mathrm{SoftDTW}_\mathrm{W2}$ & 0.69 & 0.736 & 0.737 & 0.529 \\
			\bottomrule
		\end{tabular}
	}
	\caption{Results for multi-pitch estimation on the Schubert Winterreise Dataset for SoftDTW compared with MCTC.}
	\label{tab:results:schubert1}
	\vspace*{-10pt}
\end{table}

\vspace*{-5pt}
\subsection{Incorporating Note Durations}
\label{sec:softdtw-mpe:results2}
In contrast to MCTC, SoftDTW is able to incorporate (approximate) note durations during training. SWD, for example, contains non-aligned score representations of the pieces performed. We now use these score representations as target sequences $Y$ (denoted by $\mathrm{SoftDTW}_\mathrm{W3}$, see \Fig{fig:label-types}d for an illustration). \Table{tab:results:schubert2} shows evaluation results, which are slightly improved compared to training without note durations (F-measure of $0.71$ compared to $0.69$ and $\text{CS}=0.756$ compared to $0.736$ for $\mathrm{SoftDTW}_\mathrm{W2}$). Here, there is only a moderate difference between the lengths of excerpt and label sequence and stretching the label sequence to the length of the input yields nearly identical results (denoted by $\mathrm{SoftDTW}_\mathrm{W4}$, see \Fig{fig:label-types}e). Finally, we may also use SoftDTW using strongly aligned label sequences (denoted by $\mathrm{SoftDTW}_\mathrm{S}$). In this very optimistic scenario, no alignment is necessary, but SoftDTW may compensate for inaccuracies introduced by the dataset annotation procedures. Indeed, this scenario yields best results (F-measure of $0.72$ and $\text{AP}=0.769$), even slightly improving upon the cross-entropy baseline in \Table{tab:results:schubert1}.

\begin{table}[t]
	\centering
	{\footnotesize
		\setlength{\tabcolsep}{4.5pt}
		\begin{tabular}{lcccr}			
			\toprule
			\textbf{Scenario} & \textbf{F-measure} & \textbf{CS} & \textbf{AP} & \textbf{Acc.} \\ \midrule
			$\mathrm{SoftDTW}_\mathrm{W3}$ & 0.71 & 0.756 & 0.755 & 0.552 \\
			$\mathrm{SoftDTW}_\mathrm{W4}$ & 0.71 & 0.757 & 0.750 & 0.555 \\
			$\mathrm{SoftDTW}_\mathrm{S}$ & 0.72 & 0.761 & 0.769 & 0.563 \\
			\bottomrule
		\end{tabular}
	}
	\caption{Results on the Schubert Winterreise Dataset for incorporating note durations with SoftDTW.}
	\label{tab:results:schubert2}
	\vspace*{-10pt}
\end{table}

\vspace*{-5pt}
\subsection{Cross-Dataset Experiment}
\label{sec:softdtw-mpe:results3}

\begin{table}[t]
	\centering
	{\footnotesize
		\setlength{\tabcolsep}{4.5pt}
		\begin{tabular}{lcccr}			
			\toprule
			\multirow{2}{*}{\textbf{Scenario}} & \multicolumn{4}{c}{\textbf{AP}} \\
			& \textbf{SWD} & \textbf{Bach10} & \textbf{TRIOS} & \textbf{Phenicx} \\ \midrule
			\textbf{Default network architecture} & & & &\\
			CE \cite{WeissP21_MultiPitchMCTC_WASPAA} & 0.684 & 0.864 & 0.825 & 0.829 \\
			MCTC \cite{WeissP21_MultiPitchMCTC_WASPAA} & 0.666 & 0.861 & 0.824 & 0.833 \\
			$\mathrm{SoftDTW}_\mathrm{W2}$ & 0.665 & 0.835 & 0.812 & 0.788 \\\midrule
			\textbf{Larger network architecture} & & & &\\
			CE \cite{WeissP21_MultiPitchMCTC_WASPAA} & 0.701 & 0.886 & 0.863 & 0.846 \\
			MCTC \cite{WeissP21_MultiPitchMCTC_WASPAA} & 0.677 & 0.871 & 0.849 & 0.850 \\
			$\mathrm{SoftDTW}_\mathrm{W2}$ & 0.682 & 0.896 & 0.864 & 0.838 \\
			\bottomrule
		\end{tabular}
	}
	\caption{Results for multi-pitch estimation in a cross-dataset experiment. Here, MAESTRO and MusicNet have been used for training while four different smaller datasets are used for testing.}
	\label{tab:results:others}
	\vspace*{-10pt}
\end{table}

We also perform a cross-dataset experiment (again following the setup in \cite{WeissP21_MultiPitchMCTC_WASPAA}), where we train on the popular MAESTRO \cite{HawthorneSRSHDE19_MAESTRO_ICLR} and MusicNet \cite{ThickstunHK17_MusicNet_ICLR} datasets. Both contain strongly aligned pitch annotations for the training recordings, but they do not provide non-aligned score representations of the pieces, so $\mathrm{SoftDTW}_\mathrm{W3}$ and $\mathrm{SoftDTW}_\mathrm{W4}$ are not applicable here. We then evaluate on the four smaller datasets SWD, Bach10 \cite{DuanPZ10_MultiF0_TASLP}, TRIOS \cite{FritschP13_ScoreInformedSourceSepNMFAndSynth_ICASSP} and Phenicx Anechoic \cite{MironCBGJ16_OrchestraSourceSeparation_JECE}. Note that the latter three datasets each contain less than ten minutes of audio. This is a difficult scenario since some styles and instruments in the test datasets are not present during training. For example, Phenicx Anechoic contains orchestral instruments, while MAESTRO and MusicNet contain piano and chamber music. 

The results of this experiment are given in \Table{tab:results:others}. 
Here, MCTC and a cross-entropy baseline perform roughly on par. SoftDTW yields slightly lower results, especially on Phenicx ($\text{AP}=0.788$ compared to $0.833$ for MCTC). Given that this evaluation scenario is harder and the training datasets are larger, we also repeat this experiment with a larger network architecture (increasing the number of channels for all convolutional layers in the network). The resulting architecture has roughly 600\,000 parameters, compared to 50\,000 parameters in the default architecture. Results are shown in the lower half of \Table{tab:results:others}. Average precision scores improve consistently across all methods and datasets\cegc $\text{AP}=0.896$ for SoftDTW on Bach10 compared to $0.835$ using the smaller architecture. In particular, SoftDTW now outperforms MCTC on all test datasets except for Phenicx, where the performance gap is now much smaller ($\text{AP}=0.838$ compared to $0.850$ for MCTC).

All in all, we conclude that the results for MCTC and SoftDTW are roughly comparable, even in a challenging cross-dataset evaluation. 
Thus, MCTC may be replaced with SoftDTW without sacrificing alignment quality. In addition, SoftDTW can generalize to other kinds of target sequences, as discussed in the next section.

\vspace*{-5pt}
\section{Extension to Real-Valued Targets}
\label{sec:real}
As explained in \Sect{sec:softdtw}, the two sequences $X$ and $Y$ that are used as input to SoftDTW may come from arbitrary feature spaces. In order to illustrate the potential of using SoftDTW for learning from arbitrary sequences, we now perform two experiments with real-valued targets\ciec $y_n\in\R^{72}$ for the elements $y_n$ of $Y$. Note that MCTC is unable to handle such a setting.

\vspace*{-5pt}
\subsection{Pitch Estimation with Overtone Model}
\label{sec:real:overtone-targets}
First, we consider a straightforward extension of MPE, where we transform the binary, multi-hot target vectors of MPE to real-valued vectors by adding energy according to a simple overtone model, see \Fig{fig:teaser}c. Here, we consider 10 overtones for each active pitch, with amplitude $(1/3)^n$ for the $n$-th overtone. As a baseline utilizing strongly aligned labels, we compare with a model trained using an $\ell_2$ regression loss at each frame (similar to the cross-entropy baseline in \Sect{sec:softdtw-mpe}). To evaluate, we use the cosine similarity CS between network outputs and annotations. Note that other MPE evaluation metrics are not applicable for real-valued vectors.

When performing this experiment on the SWD dataset, we obtain $\text{CS}=0.794$ for  per-frame training with strongly aligned labels, which is higher than for MPE on SWD (cf. \Table{tab:results:schubert1}). Training without strongly aligned labels using $\mathrm{SoftDTW}_\mathrm{W2}$ yields only slightly lower cosine similarities at $0.770$. This illustrates that SoftDTW also works for settings with real-valued target sequences.

\vspace*{-5pt}
\subsection{Cross-Version Training}
\label{sec:real:cross-version-training}
Second, as a scenario with more realistic target sequences, we choose $Y$ to be the CQT representation of another version (\iec a different performance) of the piece played in $X$. In this case, the two sequences $f(X)$ and $Y$ will not correspond temporally, but SoftDTW can be used to find an appropriate alignment during training. We perform this experiment using SWD, which provides multiple versions of the same musical pieces. In particular, we choose one version (OL06) as the target version and train our network using SoftDTW to align input excerpts from other versions to excerpts from OL06. Finally, we pass versions unseen during training through the trained network and evaluate against excerpts from OL06 using cosine similarity. As a learning-free baseline, we also compute CS between the original CQT representations of the test recordings and the OL06 representations. To compute the cosine similarities during testing, we use the ground truth alignments between OL06 and all other versions provided by the dataset, but we do not need ground truth alignments during training.

Directly comparing the CQT representations of input version and target yields an average cosine similarity of $0.576$. Training (using $\mathrm{SoftDTW}_\mathrm{W3}$) yields much higher results at $\text{CS}=0.720$. Thus, the network trained using SoftDTW is able to produce real-valued outputs that are similar to the target version.

\vspace*{-5pt}
\section{Conclusion}
\label{sec:conclusion}
In this paper, we have considered SoftDTW as a tool for dealing with weakly aligned learning problems in MIR, in particular, multi-pitch estimation. We showed that a network trained with SoftDTW performs on par with the same network trained using a state-of-the-art multi-label CTC loss. We further demonstrated that SoftDTW can be used to learn features when the target sequences have real-valued entries---something not possible with CTC.

In future work, SoftDTW may be applied to more diverse MIR tasks, such as lyrics alignment, audio--audio synchronization, or cross-modal learning from unaligned video--audio pairs. Furthermore, one may explore the possibility of combining both strongly aligned and non-aligned data within the same training. All these options are supported by the same algorithmic framework.

\end{document}